\documentclass{svproc}
\usepackage[utf8]{inputenc}

\usepackage{url}

\usepackage[sectionbib]{natbib}
\bibpunct{(}{)}{;}{a}{}{,}%
\usepackage{graphicx}
\usepackage{amsmath,amssymb}
\usepackage{multirow}
\usepackage{longtable}

\usepackage[b5paper]{geometry}
\begin{document}
        \mainmatter              
        \title{Evolutionary link between globular clusters and circumgalactic clouds}
        \titlerunning{Evolutionary link between GCs and CGCs}  
        %
        \author{Irina Acharova\inst{1} \and Margarita Sharina\inst{2}
        }
        \authorrunning{Irina Acharova \and Margarita Sharina} 
        %
        \tocauthor{Irina Acharova, Margarita Sharina}
        \institute{Department of Physics, Southern Federal University, \\
        5 Zorge, Rostov-on-Don, 344090, Russia.
                \email{iaacharova@sfedu.ru},\\ 
                \and Special Astrophysical Observatory, Russian Academy of Sciences, \\
                Nizhny Arkhyz, 369167, Russia. \email{sme@sao.ru}
        }

        \maketitle

\begin{abstract}
The established by us possibility to consider circumgalactic clouds (CGCs) as the remnants 
of the parent clouds in which globular clusters (GCs)
have been formed (Acharova \& Sharina 2018) is based on a comparison of the following facts. First, the metallicities of CGCs at redshifts $ z <1 $ and of GCs in our and other 
galaxies show a bimodal distribution with a minimum near $\rm [Mg/H]=-1$. Mean values and standard deviations of the Mg 
abundances in GCs and CGCs with $\rm [Mg/H]<-1$ and $\rm [Mg/H]> -1$
coincide within the typical error of measuring the elemental abundances in clouds: 0.3~dex (Acharova \& Sharina 2018). 
Second, a similar coincidence is observed for GCs and CGCs with $\rm [X/H]<-1$ and $\rm [X/H]> -1$ at redshifts 
$ 2 <z <3 $, where $[X/H]$ is the metallicity determined from the sum of several elemental abundances 
(Dias et al. 2016, Rafelski et al. 2012, Wotta et al. 2019, Quiret et al. 2016). Third, high-metallicity CGCs are observed starting from redshifts  $\rm z\le 2.5$, 
i.e. approximately 11~Gyrs ago. At the same time globular clusters were actively formed, and their
supernovae were able to enrich the surrounding gas, from which the high-metal component of the clouds was formed. 
        
\end{abstract}

\section{Observational properties of globular clusters and circumgalactic clouds}
\begin{figure*}
\scriptsize
\includegraphics[width=6cm]{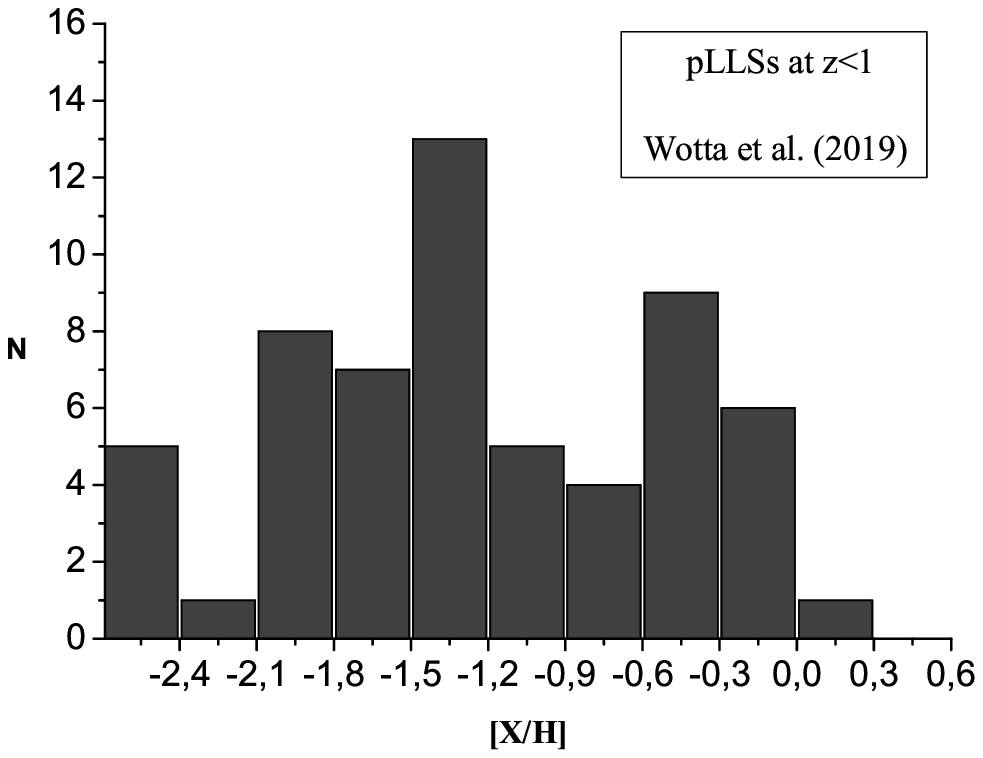}
\includegraphics[width=6cm]{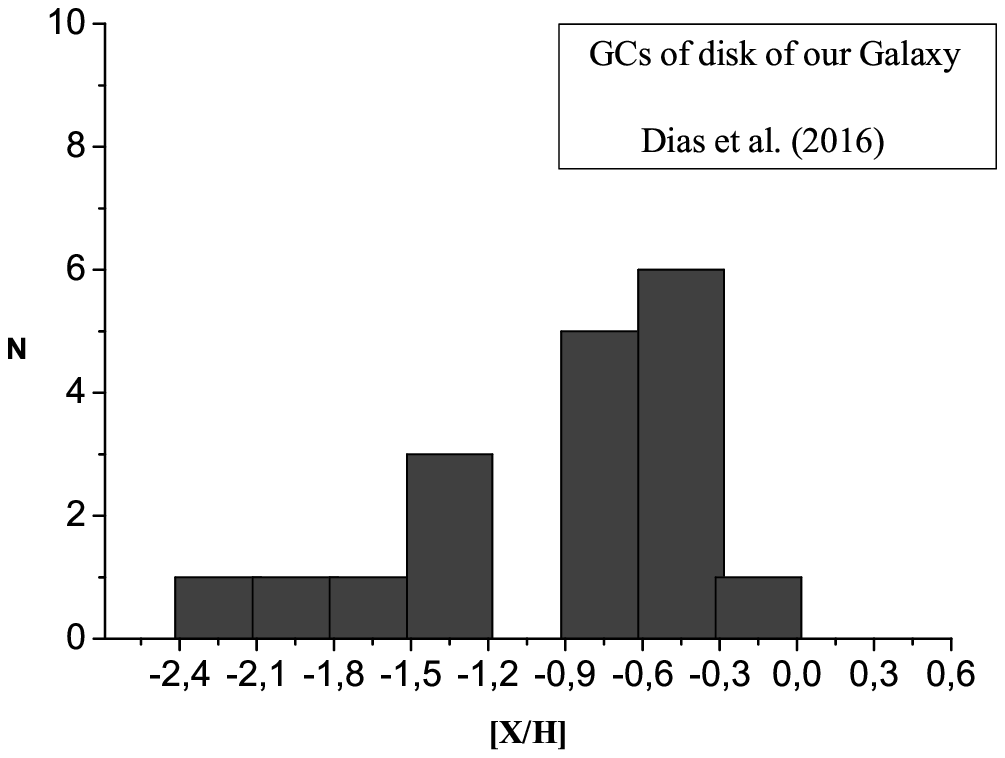}
\includegraphics[width=6cm]{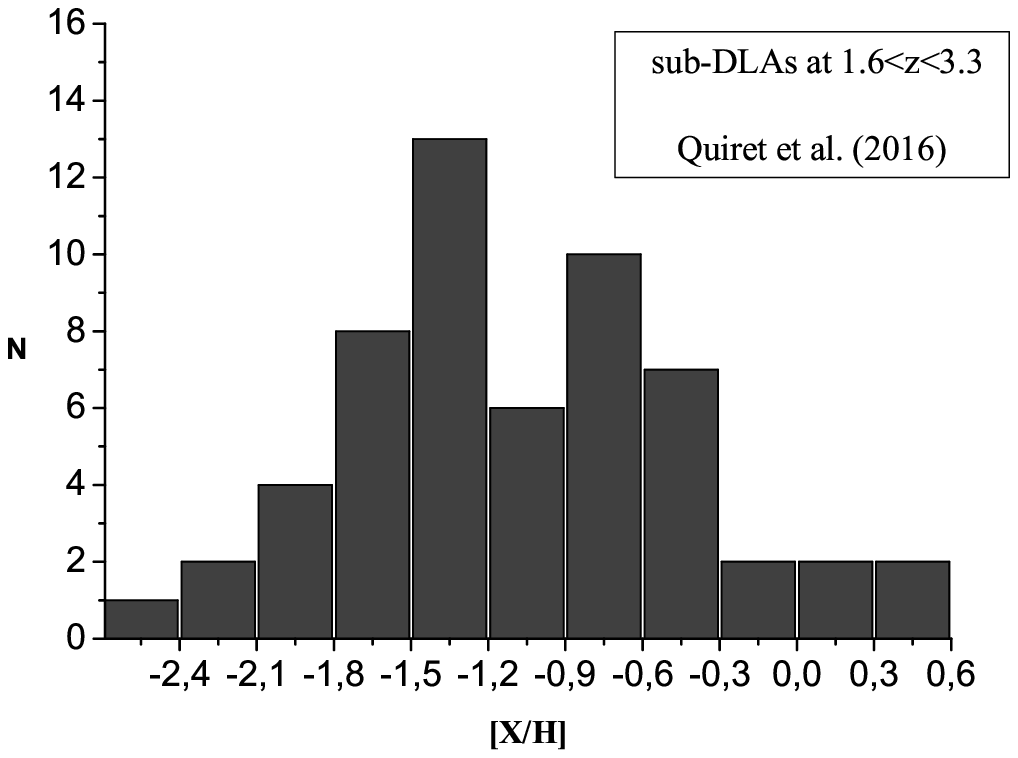}
\includegraphics[width=6cm]{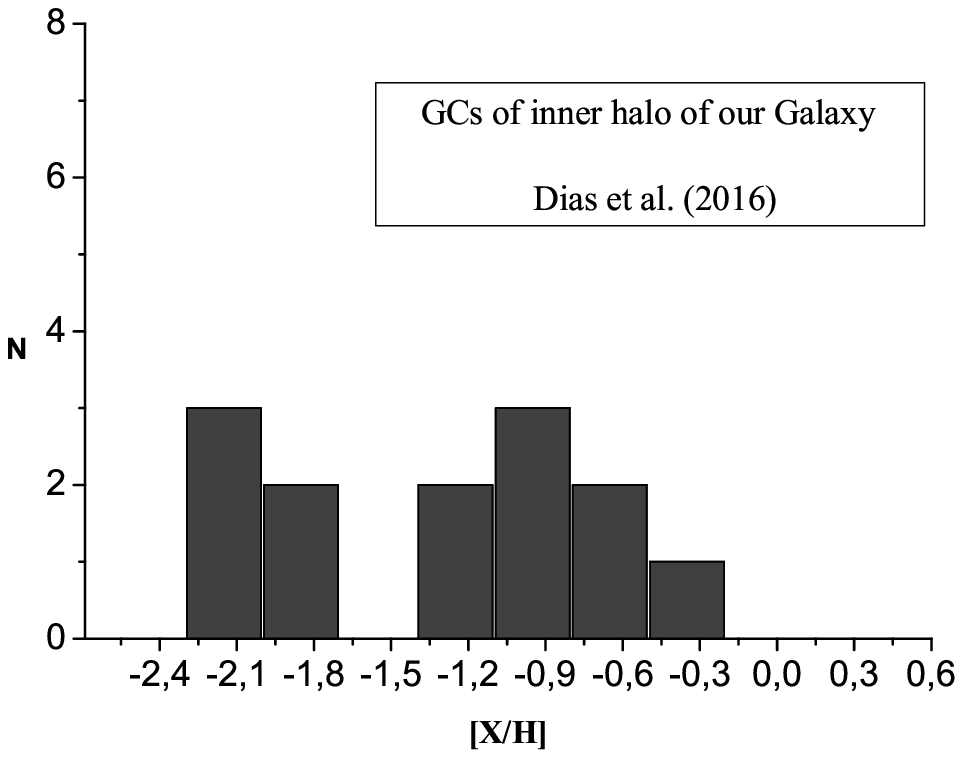}
\includegraphics[width=6cm]{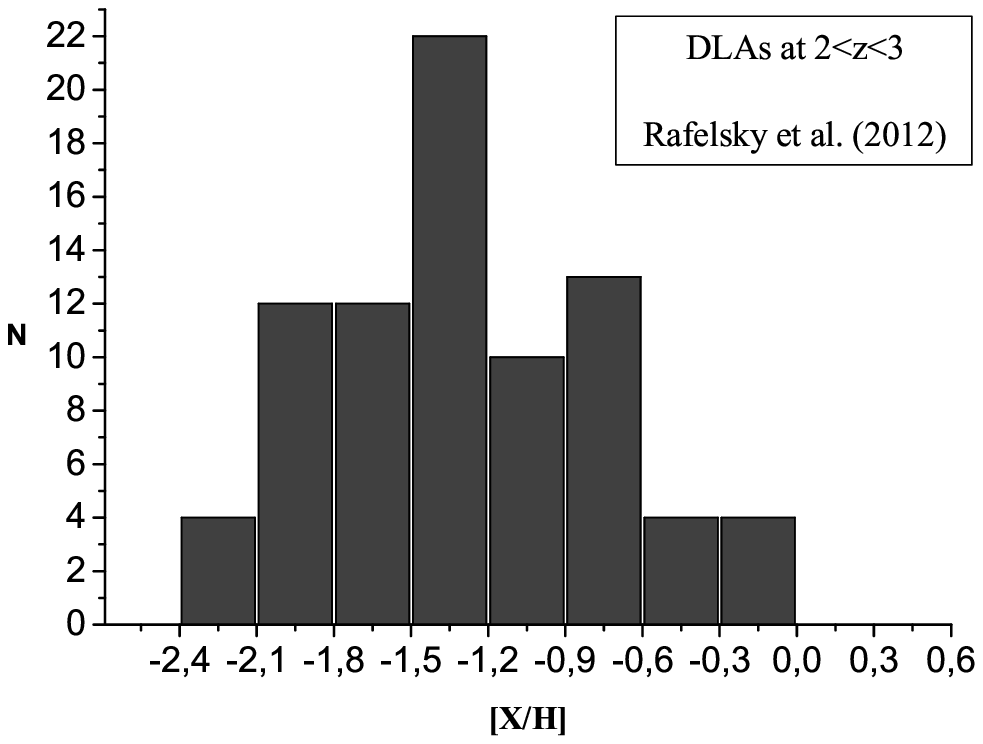}
\includegraphics[width=6cm]{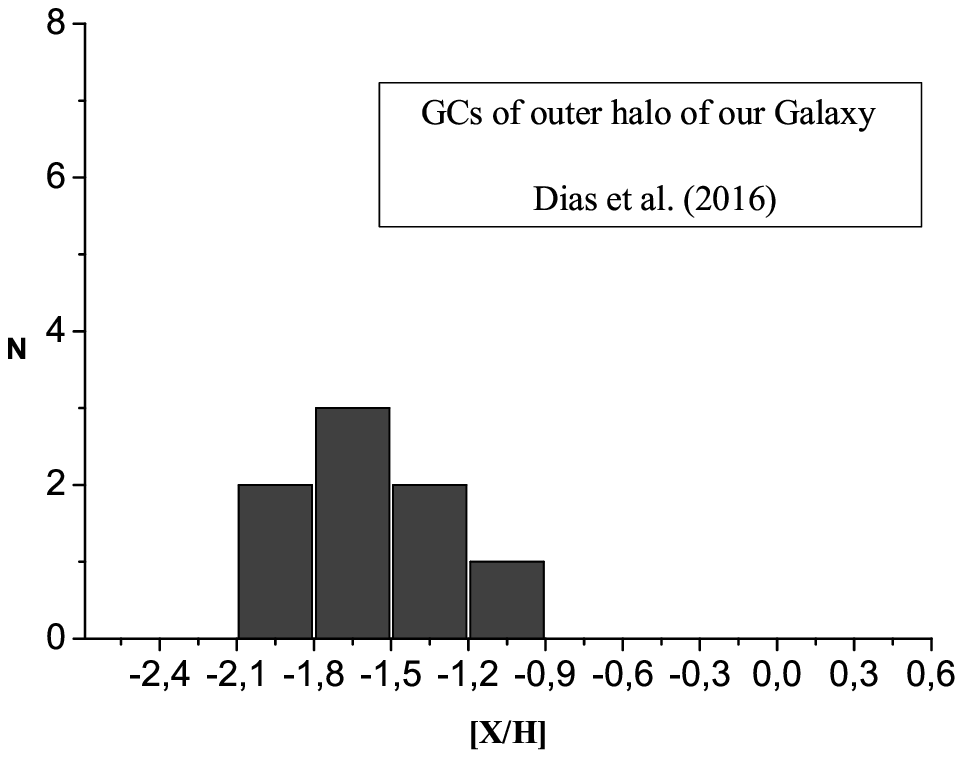}
\caption{ Left column -  metallicity distribution of
CGCs located at different redshifts. Right column -metallicity distribution of GCs in our Galaxy.}
\label{fig1}
\end{figure*} 
  \begin{table*}
\scriptsize
\caption{ Metallicities of CGCs used in the statistical analysis.}
 \begin{center}
\begin{tabular}{l|c|ccc|cccc}
\hline 
  &  &  \multicolumn{3}{c}{Low-metallicity subgroup } & \multicolumn{3}{c}{High-metallicity subgroup} &  \\ %
  &  &  \multicolumn{3}{c}{($\rm [X/H] <-1$)}           & \multicolumn{3}{c}{($\rm [X/H] \ge-1$)}                 &    \\ \hline

objects &  redshifts  & $\rm <[X/H]> $ & $\rm \sigma([X/H])$ & number & $\rm <[X/H]>$ & $\rm \sigma([X/H])$ & number & Ref\\ 
   & &           &      &  of clouds  &         &        &  of clouds    &     \\ \hline

DLAs & $2<z<3$&-1.57 & 0.35 & 57 & -0.62 & 0.22 & 24 & [7] \\
sub-DLAs & $1.6<z<3.3$ &-1.52 & 0.35 & 33 & -0.44 & 0.39 & 24 & [6] \\
DLAs & $z<1$ & -- & -- & -- & -0.57 & 0.26 & 29 & [9] \\
pLLSs & $z<1$ & -1.75 & 0.46 & 37 & -0.50 & 0.27 & 23 & [9] \\
LLSs & $z<1$ & -1.60 & 0.42 & 11 & -0.45 & 0.19 & 9 & [9] \\
LLSs & $2.3<z<3$ & -1.69 & 0.20 & 8 & -- & -- & -- & [4] \\
\hline
\end{tabular}
 \end{center}
\end{table*}
The study compares the abundance properties of circumgalactic clouds (CGCs) that are detected by studying absorption lines in spectra of quasars. 
CGCs are classified according to their neutral hydrogen column
densities $\rm logN_{HI}$ as follows: damped Lyman limit systems (DLAs) with $\rm logN_{HI} \ge 20.3$, sub-damped 
Lyman limit systems (sub-DLAs) with $ \rm 19 \le logN_{HI} < 20.2$, Lyman limit systems (LLSs) with $\rm 17.2 \le logN_{HI}<19$, 
and partial Lyman limit systems (pLLSs) with $\rm 16.1 \le logN_{HI} < 17.2$. The lower $\rm logN_{HI}$ , 
the higher the gas ionization rate is. At redshifts $ z <1 $, all the studied clouds are located within 100~kpc from galaxies (Lehner et al., 2013) , 
i.e. belong to their circumgalactic medium. At high redshifts $ \rm z \sim 3 $, all the clouds are located within virial radii of galaxies (Rudie et al., 2012).
As it was stressed in the abstract, there are several observational facts, the comparison of which allows us to definitely
consider CGCs as the remnants of the parent clouds in which globular clusters (GCs) have been formed.

Figure~1 illustrates the metallicity distributions of CGCs located at different redshifts (left column) and of GCs of our Galaxy (right column).
It can be seen in the first row of Fig.~1 that the distribution of CGCs  (Wotta et al., 2019) looks similar to the distribution of GCs in the disk of our Galaxy (Dias et al., 2016).
The second row of Fig.~1 illustrates the similarity of the distributions of CGCs (Quiret et al., 2016) and GCs in the inner Galactic halo (Dias et al., 2016).
The distribution of CGCs from the paper by Rafelski et al. (2012) and GCs in the outer Galactic halo (Dias et al., 2016) are shown in the third row of Fig.~1. 
The last two distributions look similar.
Tables~1 and 2 summarize mean metallicities and the corresponding statistical dispersions for low- and high-metallicity subgroups of CGCs at various redshifts 
(reference papers are listed in the last column) 
and for low- and high-metallicity subgroups of GCs in three Galactic subsystems from Dias et al. (2016). 
\begin{table*}
 \caption{Metallicities ($\rm [X/H]$) of GCs according to Dias et al. (2016)}
 \begin{center}
\begin{tabular}{l|ccc|ccc}
\hline 
 &  \multicolumn{3}{c}{$\rm [X/H] <-1$ } & \multicolumn{3}{c}{$\rm [X/H] \ge-1$} \\  \hline

   Galactic subsystem     & $\rm \langle [X/H] \rangle$ & $\rm \sigma([X/H])$ & N$_{GCs}$& $\rm \langle[X/H]\rangle $& $\rm \sigma([X/H])$ & N$_{GCs}$ \\ \hline
disc & -1.65 & 0.30 & 6 & -0.59 & 0.17 & 12 \\
inner halo & -1.74 & 0.43 & 8 & -0.69 & 0.24 & 5 \\
outer halo & -1.62 & 0.28 & 8 & -- & -- & -- \\
\hline 
\end{tabular}
\end{center}
\end{table*}    

\section{Conclusions}

In our opinion, the following process was realized at the early stages of the galactic evolution. 
Low-metallicity CGCs were enriched with metals formed as a result of nucleosynthesis in first stars, that is, in the stars of population~III.
The mean age of metal-rich Galactic and M31 GCs is 10.637$ \pm $0.468 Gyr, and the mean age of metal-poor Galactic and M31 GCs is 10.217$ \pm $0.226 Gyr (Chattopadhyay et al., 2012).
That is, the mean ages of the groups of metal-rich and metal-poor GCs in two galaxies are nearly the same.
This means that the enrichment in metals of low-metallicity CGCs, progenitors of GCs, occurred quickly as a result of supernova explosions. 
Three types of massive stars have short lifetimes. These are supernovae type II (SNe~II), binary systems leading to explosions of SNe types Ib/c
and SNe~Ia from short-lived progenitors. There is no a generally accepted model for the last type. According to the theory of nucleosynthesis in supernovae, 
SNe~Ia are the main suppliers of iron in the circumgalactic medium.
Therefore, GCs are unique laboratories in which Ia SNe from short-lived progenitors can be studied.

\paragraph{Acknowledgments.}
 The work is supported by the Russian Foundation for Basic Research (grant no. RFBR 18-02-00167 a).

%
%

\end{document}